\documentclass[conference]{IEEEtran}
\IEEEoverridecommandlockouts

\usepackage{cite}
\usepackage{amsmath,amssymb,amsfonts}
\usepackage{algorithmic}
\usepackage{graphicx}
\usepackage{textcomp}
\usepackage{xcolor}
\def\BibTeX{{\rm B\kern-.05em{\sc i\kern-.025em b}\kern-.08em
    T\kern-.1667em\lower.7ex\hbox{E}\kern-.125emX}}

\usepackage{multirow} 
\usepackage{pifont}
\usepackage{booktabs}
\usepackage{tabularx}

\usepackage[breaklinks]{hyperref}
\hypersetup{
    colorlinks=true,
    citecolor=blue,
    linkcolor =	red,
    filecolor=magenta,      
    urlcolor=cyan,
}

\begin{document}

\title{Evaluation of Machine Learning Models in Student Academic Performance Prediction}

\author{\IEEEauthorblockN{A.G.R. Sandeepa}
\IEEEauthorblockA{\textit{Dept. of Computer Engineering} \\
\textit{General Sir John Kotelawala Defence University}\\
Ratmalana, Sri Lanka \\
sandeepa.a@kdu.ac.lk}
\and
\IEEEauthorblockN{Sanka Mohottala}
\IEEEauthorblockA{\textit{Dept. of Computer Science} \\
\textit{Sri Lanka Institute of Information Technology}\\
Malabe, Sri lanka \\
sanka.m@sliit.lk}
}
\maketitle
\IEEEpubidadjcol
\begin{abstract}
This research investigates the use of machine learning methods to forecast students' academic performance in a school setting. Students' data with behavioral, academic, and demographic details were used in implementations with standard classical machine learning models including multi-layer perceptron classifier (MLPC). MLPC obtained 86.46\% maximum accuracy for test set across all implementations. Under 10-fold cross validation, MLPC obtained 79.58\% average accuracy for test set while for train set, it was 99.65\%. MLP's better performance over other machine learning models strongly suggest the potential use of neural networks as data-efficient models. Feature selection approach played a crucial role in improving the performance and multiple evaluation approaches were used in order to compare with existing literature. Explainable machine learning methods were utilized to demystify the black box models and to validate the feature selection approach. 

 
\end{abstract}


\section{Introduction}
\label{section:intro}
Currently, high schools have access to an abundance of structured and unstructured data, ranging from exam results and student registration records to attendance, involvement in educational activities, and usage of online resources. For educational institutions, this is both a chance and a problem in terms of efficiently utilizing the data for practical insights. Predicting a student's academic success is one of the most useful uses since it gives schools the ability to make better decisions and achieve better results for their students.

Machine learning has emerged as an invaluable tool for developing predictive models in a variety of fields, including education. Through the process of analyzing past data and discovering trends, machine learning models are capable of accurately predicting future student outcomes. Machine learning models have shown to be highly successful when it comes to educational data mining based tasks such as grade prediction, dropout detection, and tailored learning suggestions.

Factors that affect student academic performance prediction (SAPP) can be categorized under four areas. Demographic factors including gender, country of birth and income can have a significant impact on the students performance. Although some studies predict student performance using demographic data, these factors might not always be enough to produce reliable predictions~\cite{abu2019factors}. Behavioral factors are another dimension that need to be considered. Involvement in discussions and utilization of online learning materials are examples of behavioral characteristics that have been shown to be reliable predictors of academic success~\cite{beltran2021analysis}. Parental involvement is also another factor that affect the students performance and subsequently the predictability of student's academic performance. Perhaps the most relevant predictors of student success are academic variables, such as prior performance, test results, and grades in particular courses. Effective evaluation of academic performance should consider all these four factors~\cite{amrieh2016mining}.


When it comes to SAPP research, several datasets have been used. The Open University Learning Analytics dataset~\cite{Kuzilek2017} contains 32k records which includes demographic details as well. KDDcup~\cite{WinNT} dataset is a much larger dataset that contains data across 39 courses. HMedx~\cite{sustainability} is one of the largest datasets in SAPP domain containing 600k records. But typical data collections result in small number of records thus predictive models based on large datasets may not be useful. Furthermore, techniques developed for these datasets may not generalize well to small datasets. SAPData~\cite{amrieh2016mining} is a small dataset (480 records) containing features covering all four factors detailed above. SAPData dataset was used in our research since we expect our research to advance data collection based research in high school settings.



Based on the literature review and the gaps identified, in this paper we provide the following main contributions,  
\begin{itemize}
\item To the best of our knowledge, we are the first to do implementations across multiple ML models with hyper-parameter tuning using SAPData dataset under multiple evaluation protocols.
\item We introduce a feature selection method based on exploratory data analysis and statistical tests.
\item We show that despite the small size of the SAPData dataset, MLP outperforms other machine learning models, demonstrating the potential of neural networks as data-efficient approaches for tabular data. 
\item We employed an Explainable ML method to validate our feature selection method and to interpret the results. 
\end{itemize}
%

The rest of the paper is organized as follows. Section~\ref{section:lit_review} describes related work done in previous research. Data pre-processing, data analysis methods, experimental setup and machine learning models are described in Section~\ref{section:methodology}. Section~\ref{sec:Results_and_discussion} describes the obtained results and interpretation of those results. Finally, Section~\ref{sec:conclusion} concludes with future research directions.

\section{Literature Review}
\label{section:lit_review}
This section details the literature review done on the previous work based on SAPData dataset, machine learning use in education domain as well as on other related areas.


\subsection{Overview of Data-Driven Methods in Education }
\label{subsec:data_driven_methods_in_edu}

High school system data, such as attendance logs, assignment meta-data, and test results, offer a wealth of information for creating models that predict a student's success or failure. Use of this data to obtain useful insights is called Educational Data Mining (EDM)~\cite{amrieh2015mining}. Machine learning algorithms are extensively utilized in the mining of educational data, enabling the prediction of multiple outcomes, including student grades, probability of dropout, and overall academic accomplishment. 

Machine learning models such as Decision Trees (DT), Naïve Bayes (NB), Support Vector Machines (SVM),k-nearest neighbour (k-NN), Logistic Regression (LR) and Neural Networks (NN) have been used in past SAPP research. 
Decision trees are widely used because of their ease of interpretation and simplicity. 
Stronger alternatives for forecasting student performance are provided by SVM and Neural Networks.

\subsection{Feature selection methods used with SAPData Dataset}
\label{subsec:feature_selection}
In the study~\cite{amrieh2016mining}, the authors used the filter-method to rank features in order to determine which one is of most importance while creating a model for student performance and it is based on an information gain-based selection algorithm. To improve the effectiveness of the Modified K-Nearest Neighbor (M-KNN) algorithm in predicting students' academic performances, feature selection based on Genetic Algorithm (GA) is applied in the study~\cite{wafi2019automatic}. In the research~\cite{Tanabe}, feature selection has been carried out using a way that the features are converted to binary representations and a machine learning model is used to assess their relevance. 

\subsection{Machine Learning Models using SAPData Dataset }
\label{subsec:ML_in_edu}

Earlier version of the SAPData was used in~\cite{amrieh2015mining} with 150 records. Five-fold cross validation (CV) was used for evaluation which resulted in 73.8\% accuracy with an MLP classifier. Current version of SAPData with 500 records~\cite{amrieh2016mining} used information gain based feature selection method to select the 10 best features. Ten-fold CV was used in evaluations with Decision Trees (DT), Naive Bayes (NB), Random Forest (RF) and MLP models. Bagging and Boosting were also used and obtained 79.1\% for MLP classifier. Furthermore, implementations were done with dataset containing all features (called WBF) as well as with all features except behavioral features (called WOBF).

Multiple ML models including DT, NB, MLP,k-NN were used~\cite{rahman2017predict} and obtained 78\% accuracy with MLP using all features (WBF). With the use of an ensemble filtering method to select data, they improved their accuracy to 84\%. In~\cite{wafi2019automatic}, k-NN was used with a genetic algorithm which obtained 82\% accuracy. But vanilla k-NN resulted in 73\% accuracy. All feature were used in both of those implementations and evaluation method (k-fold CV etc.) was not mentioned. Francis and Babu~\cite{francis2019predicting} obtained 75\% for DT implementation and 10-fold CV was used with all features in this implementation. Most of their other implementations resulted in less than 68\% accuracy. Roy and Farid~\cite{roy2024adaptive} introduced  recursive feature selection method and forward selection method and obtained 78\% with logistic regression model, 81\% with SVM.

\subsection{Explainable AI in Education}
\label{subsec:Explainable_AI}
The study~\cite{Lundberg} provides SHapley Additive exPlanations (SHAP), a unified framework for prediction interpretation, as a solution to interpret the predictions of complex models. Researchers~\cite{wang} proposes the XGB-SHAP model, a method for predicting student achievement that combines SHAP with Extreme Gradient Boosting (XGBoost).To engage faculty members and advisers in the creation of an interactive knowledge discovery tool for a better understanding of student achievement and students at risk, the study~\cite{Nur} uses explainable AI tools such as LIME and SHAP in a human-centered design approach. 
Hayat et al~\cite{hayat} employs SHAP values on SAPData dataset to explain how ensemble algorithms predict student performance.

\section{Methodology}
\label{section:methodology}

\subsection{Exploratory Data Analysis}
\label{subsection:dataset}

SAPData~\cite{amrieh2016mining} dataset contains data from Jordan high school students which were collected using a learning management system (LMS) called Kalboard 360. Unlike other EDM datasets that capture details of students from single grade or performance for a single subject, SAPData contains details across multiple grades and multiple subjects. 
SAPData contains 480 records with 16 features capturing multiple areas that can potentially indicate the students' academic performance and those details are given in the Table~\ref{T:SAPData_features}. As for the label, we use the discrete marks of each student for the subject for which the record was collected. If the marks are less than 70, then it is categorized as Low Level ('L'), if it is between 70-89 then to Middle Level ('M') and if it is above 90, then to the High Level ('H'). Since categorical features of the dataset contain strings values, unique values of each categorical variable were identified and with the use of Label Encoding, these values of the variables were converted to positive integer values.

\begin{figure}[bp]
    \centering
    \includegraphics[width=\linewidth]{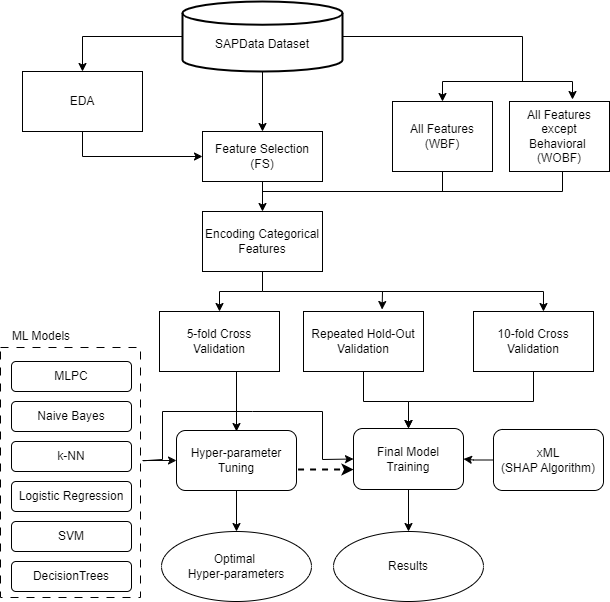}
    \caption{Overview of the Methodology}
    \label{fig:block_diagram}
    \vspace{-3mm}
\end{figure}

Identification of feature type along with summary statistics were done as the first step, after which visualizations were done to get a better understanding. Scatter plot across all features was used to visualize the independence of features as well as the correlation of each feature with the class value. For each categorical feature, distribution of data was visualized using bar plots. Furthermore, to understand the effect of feature value on label, bar plots were done with respect to label value as well. For numerical features, histogram plots were created to go beyond the summary statistics of these features and to get a better understanding of the distribution. Scatter plots were used to observe the effect of multicollinearity. This was further improved by observing the scatter plots under each class value. Furthermore, the Pearson correlation matrix of numerical features was also obtained. Numerical features were grouped by features such as 'Topic' and 'GradeID' to observe if there is any explainable pattern in the data as well.  

\subsection{Feature Protocols}
\label{subsection:feature_protocolsl}


Given that the dataset contains both categorical and numerical features, we introduced a feature selection method (FS) by considering these two types of features in an independent manner. Since all features, except one categorical feature are nominal, one-hot encoding is the best encoding approach but just as in EDA, we opted for label encoding.This is because, since the size of the dataset is small, one-hot encoding will result in a large number of features which can result in sub-optimal feature selection. Moreover, as detailed in Section~\ref{section:lit_review}, previous researchers also have opted for this approach. 

For categorical feature selection, experiments with two statistical methods, namely, ANOVA F-value test and chi-squared test were done. Since target variable is a categorical variable, chi-squared test was used in the final feature selection approach. For numerical feature selection, results from EDA were used. 

Apart from this protocol, full feature set based protocol named 'With Behavioral Features' (WBF) was introduced to compare the performance of our FS protocol. Since many previous research (Section~\ref{section:lit_review}) had done implementations with all features except the behavioral features, we introduce that as another protocol named 'Without Behavioral Features' (WOBF). All these 3 feature protocols (FS, WBF, WOBF) were used under multiple evaluation protocols as detailed in Section~\ref{subsection:model_eval}.   

\subsection{Data Pre-Processing}
\label{subsection:data_pre_process}
For the use with ML models, ordinal categorical features were encoded using label encoding and the nominal categorical features were encoded using one-hot encoding. For the label (target) variable, since there were 3 classes, -1 was given for 'L' and 0 for 'M' and +1 for 'H'. Preliminary experiments were done without scaling, with scaling under standardization and with scaling under normalization. Based on the results, Standardization was used for scaling in all subsequent experiments. Since we are not using a Hold-Out evaluation approach (i.e., single train-test split approach), it is not possible to do the normalization before the model training process if we want to avoid any information leakages during the standardization process. Thus, data scaling was done as part of the model training process.


\subsection{Machine Leaning Models}
\label{subsection:ml_models}
In order to do comprehensive model implementations, we selected 5 classical ML models and a neural network based model. Since we are comparing the performance of vanilla ML models, we chose not to use and bagging, ensemble learning or boosting based models. We used k-nearest neighbour (k-NN) model as our baseline model. Decision trees (DT) and naive bayes (NB) models were selected for comparison since most of the previous research had used these models. We also included soft margin Support vector machine (SVM) and the logistic regression (LR) classifiers in implementations. For NB, multi-nominal Naive Bayes was used without the feature standardization since it requires the features to be discrete values. Since this is a multi-class classification, one-vs-one was used to obtain the decision boundary for LR, SVM classifiers.

For the Neural Network (NN) model, we use a multi-layer perceptron (MLP) classifier since the MLP is a highly expressive model and since there are no known inductive biases associated with tabular datasets in general. Furthermore, to validate the potential use of NNs as a data-efficient model with a small tabular dataset, MLP is the starting point. Use of advanced neural networks such as convolutional neural network (CNN) may result in sub-optimal results since assumption of locality and translation invariance may not be present in the given tabular dataset. 

Initially, Hyper-parameter tuning was done with all models under the FS protocol and 5-fold cross validation approach and subsequent implementations used these hyper-parameters. With MLP model, we experimented only with pyramid structured architecture and only up to 5 layers with maximum of 256 nodes in a layer. Hyper-parameter search spaces for each model are given in Table~\ref{T:hyper_tuning_res}.  

\subsection{Model Evaluation Methods}
\label{subsection:model_eval}
We used different methods for dataset splits for training and validating and different metrics to evaluate the performance.

\subsubsection{Dataset Evaluation Protocols}
\label{subsubsection:mdataset_eval}
As we observed in Section~\ref{subsec:ML_in_edu}, there is no standard evaluation method used with SAPData dataset. In order to compare with previous research benchmark results, we used three evaluation approaches. Stratified shuffle was used in all data splitting approaches to reduce the adverse effects of data imbalance. 
\begin{itemize}
     \item Repeated Hold-Out Validation: randomly split the dataset 90:10 for training and testing the model and repeat this process for 100 different times. Reduces bias of dataset and result in model stability.  
     \item 5-fold Cross Validation: Cross validation with 5 folds used for hyper-parameter tuning.
     \item 10-fold Cross Validation: cross validation with 10-folds used to evaluate the model but repeated the process for 10 times. 
   \end{itemize}


\subsubsection{Metrics for Performance Evaluation}
\label{subsubsection:model_eval}
Average accuracy, maximum accuracy and standard deviation of accuracy results were reported for 10-fold CV and RHO while for 5-fold CV, average accuracy was reported. Confusion matrices were used with RHO along with precision, recall and f1-score values. Box plots were used for comparison of different models.

\subsection{Explainable Machine Learning }
\label{subsection:xML}
SHAP (SHapley Additive exPlanations) algorithm was used as a method to explain the black box models such as MLP, LR and SVM. For a given datapoint, for each feature there is an associated SHAP value and if it  is positive, then it positively contribute to the prediction and if it is negative it contributes negatively. Summary plots of SHAP values detail the SHAP value distribution for all samples and we use those plots to explain the trained models.

\subsection{Experimentation}
\label{subsection:experiments}
For experiments, we used jupyter notebook environment and for all models including the MLP, we used sklearn library. Implementation codes will be released upon request\footnote{Request by filling the \href{https://forms.gle/wdJiHnGb1yJXPfWs7}{form}}.


\section{Results and Discussion}
\label{sec:Results_and_discussion}

\subsection{Preliminary Results}
\label{subsec:eda_res}

\begin{figure}[tbp]
    \centering
    \includegraphics[width=\linewidth]{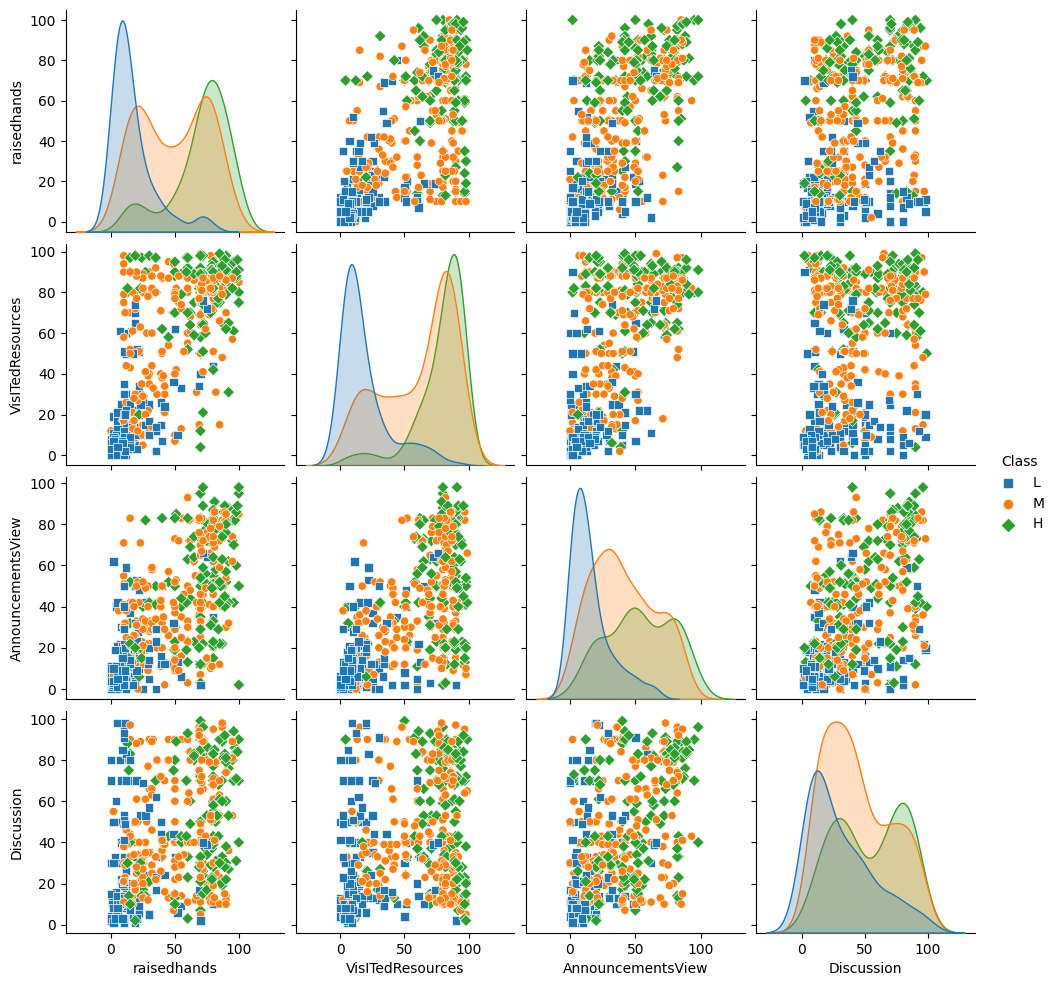}
    \caption{Scatter plots for behavioral features}
    \label{fig:behavior_scatter}
    \vspace{-6mm}
\end{figure}

Dataset is an imbalanced dataset where 26.46\% of dataset was for 'L' label, 43.96\% for 'M' label and 29.58\% for 'H' label. Behavioral features ranged within 0-100 in distributions (Figure~\ref{fig:behavior_scatter}) and the observable left skewed and right skewed behaviors are respectively associated with High marks and Low marks. Scatter plots further show that there is very little correlation between these 4 numerical variables. Thus all behavior features were included in the FS protocol. 

Both ANOVA F-test and chi-squared test resulted in lowest ranking for \{'SectionID', 'StageID', 'GradeID', 'Semester'\} features. For chi-squared test, we used 5\% as significant value and all above features except 'GradeID' resulted in p-values that are above 5\%. Since 'GradeID' was the next in-line, we removed all those 4 categorical features and selected others. Under One-hot-encoding, final datasets resulted in 63 features for WBF protocol, 59 features for WOBF protocol and finally 54 features for SF protocol.  

\subsection{Machine Learning Model Implementations}
\label{subsection:ml_model_res}

\begin{table}[bp]
\centering
\caption{Hyper-parameter tuning results}
\label{T:hyper_tuning_res}
\resizebox{\columnwidth}{!}{
\begin{tabular}{clll}
\hline
Model & Hyper-parameter & Search Space & Selected Parameters  \\
\hline
\addlinespace[1pt]

SVM         &  C      & 100, 10, 1, 0.99, 0.9, 0.4, 0.1        & 1        \\\hline  
\multirow{4}{*}{\centering MLP}              & layer architecture  & maximum of 256 nodes and 5 layers          & [128,64,32,16,8]            \\
          & activation  &relu, tanh         & relu \\
         & batch size  & 100, 200         &100 \\ 
          & learning rate  &0.0001, 0.001, 0.1    & 0.0001 \\
 \hline  
\multirow{2}{*}{\centering k-NN}       & k      & 3, 5, 7, 9, 11, 13, 15, 17, 19  & 15            \\
          & p  &1, 2, 3, 4, 5         & 1 \\ \hline
\multirow{2}{*}{\centering LR}       & Regularization      & L1, L2  & L2            \\
          & C  &100, 10, 1, 0.5, 0.1         & 0.1 \\ \hline
NB &alpha &1, 0.5, 0.1  &1 \\ \hline
\multirow{5}{*}{\centering DT}  & criterion  & gini, entropy          & gini            \\
          & splitter  & random, best         & random \\
         & maximum depth  & None, 2, 3, 4, 5, 6, 7, 8, 9, 10, 20          &6 \\ 
          & max samples per split  & 2, 4, 8, 16    & 4 \\
          & max samples per leaf  &1, 2, 4, 8   & 4 \\ \hline
\end{tabular}%
}
\vspace{-7mm}
\end{table}


\vspace{-1.5mm}
Initial model implementations were done to find the hyper-parameters using grid search approach under 5-fold CV using the SF protocol. Search space of those hyper-parameters and the resultant optimal hyper-parameters are given in the Table~\ref{T:hyper_tuning_res}. The C parameter controls the strength of L2 regularization in SVM and strength of L1/L2 regularization in LR. Average test accuracy for all models are given in the table~\ref{T:ml_model_acc} under 5-fold CV column. Contrary to the results from other evaluation protocols, drop of accuracy was noticed for all models except DT. Test accuracy distribution of each split suggest this could be due to the bias in the data splits used in 5-fold CV.

\begin{figure}[tbp]
    \centering
    \includegraphics[width=\linewidth]{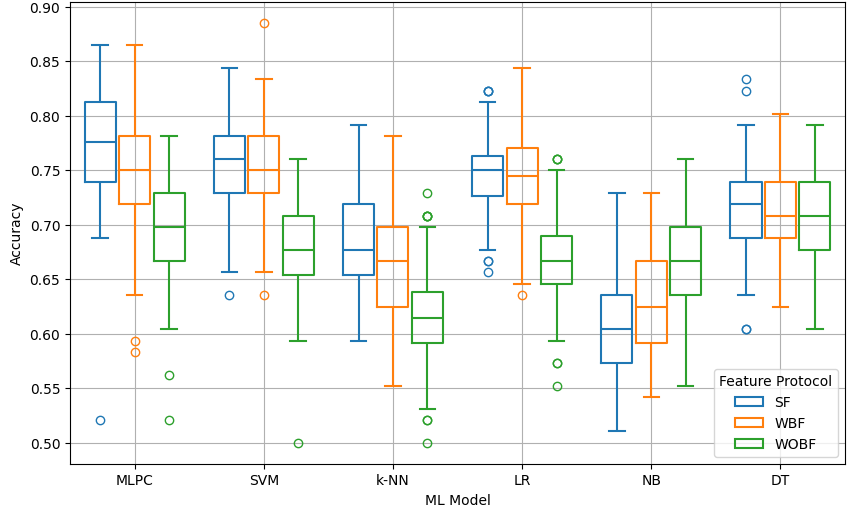}
    \caption{Repeated Hold-Out (RHO) validation results for all ML models}
    \label{fig:rho_box_plot}
    \vspace{-3mm}
\end{figure}

Average accuracy results (Table~\ref{T:ml_model_acc}) and the accuracy distribution results (Figure~\ref{fig:rho_box_plot}) under RHO protocol suggest that all models except NB result in high accuracy with SF protocol compared to WBF protocol, and WBF result in higher accuracy compared to WOBF. Similar results can be observed with 10-fold CV to a lesser degree as well. This validates the superiority of our feature selection approach and the importance of behavioral features in model predictions. Since the naive baseline of the dataset is 37.08\%, results obtain for all models show considerable improvement over this. Standard deviation values, box plot distributions and existence of outliers suggest that there is some variation in all models which could be attributed to composition and the small size of the dataset. NB performance is the opposite of other models (Figure~\ref{fig:rho_box_plot}), which can be attributed to the assumptions of the Multi-nominal NB classifier. It assumes features are categorical features but existence of numerical features in SF and WBF can be attributed to the lower performance of them compared to WOBF. Better performance of WBF over SF can be attributed to the existence of more categorical features in WBF compared to SF. Compared to the other models, DT models show a stable performance across all evaluation protocols which could indicate either less overfitting effect or better hyper-parameter selection compared to other models.        

\begin{figure}[tbp]
    \centering
    \includegraphics[scale=0.5]{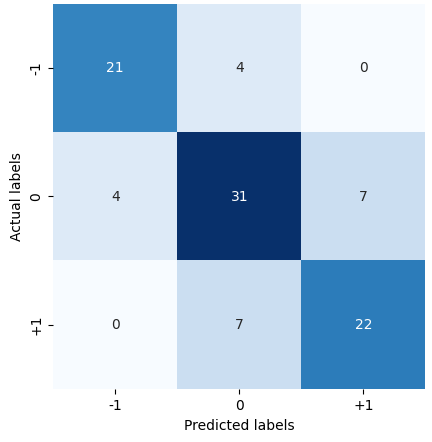}
    \caption{Confusion matrix for MLP (under SF / RHO protocols)}
    \label{fig:conf_mlp_sf_rho}
    \vspace{-3mm}
\end{figure}

Results further indicate that average accuracy of MLP is the best across both the 10-fold CV and RHO protocols. Furthermore, under 10-fold CV, maximum accuracy out of the 10 repetitions result in the best accuracy for MLP as well (Table~\ref{T:ml_model_acc}). Furthermore, under RHO, MLP result in the 2nd best accuracy after SVM. Moreover, when considering under each feature protocol under RHO, MLP result in best accuracy for SF while it result in 2nd best under WBF and WOBF (Fig.~\ref{fig:rho_box_plot}). When considering the MLP model predictions for all test samples (Figure~\ref{fig:conf_mlp_sf_rho}), we observe these is little to no misclassifications of 'H' samples as 'L' and vice-versa. On the other hand, for 'M' samples, misclassifications as 'L' and 'H' are prevalent. This suggests that MLP performance is tightly regulated by the 'M' samples. These overall results indicate that even though neural networks generally perform poorly with tabular data and with small datasets, they have the potential to act as data-efficient models. 

\begin{table}[htbp]
\centering
\caption{ML Model Performance}
\label{T:ml_model_acc}
\resizebox{\columnwidth}{!}{%
\begin{tabular}{ccclllllll}
\hline 
\addlinespace[1pt]
 {\multirow{2}{*}{\rotatebox[origin=c]{0}{Model}}}   & \multicolumn{1}{c}{5-fold CV} & & \multicolumn{3}{c}{10-fold CV (n=10)} & & \multicolumn{3}{c}{RHO (n=100)}   \\ 
 \cmidrule{2-2} \cmidrule{4-6} \cmidrule{8-10} 
    & SF      &    & SF    &WBF   &WOBF &  &SF   &WBF   &WOBF \\ \hline \addlinespace[1pt]
\multirow{2}{*}{\centering MLPC}   & \multirow{2}{*}{\centering 61.04}  &          & 79.58 &76.56 & 69.58  &              & 77.25  &74.61 &69.67  \\ \addlinespace[-2pt] 
   &   &                                                                           & (80.00) &(78.12) & (70.83)  &                                  & (4.91)  &(5.10) &(4.76)  \\ \addlinespace[2pt]
   
\multirow{2}{*}{\centering SVM}   & \multirow{2}{*}{\centering 61.87}   &          & 75.89 &76.06 & 68.43  &              & 75.42  &75.20 &67.45  \\ \addlinespace[-2pt] 
   &   &                                                                            & (77.70) &(77.29) & (70.20)  &                               & (3.95)  &(4.07) &(4.11)  \\ \addlinespace[2pt] 
   
\multirow{2}{*}{\centering k-NN}   & \multirow{2}{*}{\centering 63.33}   &      & 67,79 &65.91 & 62.31  &          & 68.32  &66.04 &61.37  \\ \addlinespace[-2pt] 
   &   &                                                                        & (70.0) &(67.08) & (63.75)  &                                   & (4.51)  &(4.49) &(4.38)  \\ \addlinespace[2pt] 
   
\multirow{2}{*}{\centering LR}   & \multirow{2}{*}{\centering 63.12}   &            & 75.0 &74.77 & 67.06   &           & 74.57  &74.34 &66.68  \\ \addlinespace[-2pt] 
   &   &                                                                            & (76.87) &(75.83) & (69.58)  &                           & (3.58)  &(3.77) &(4.10)  \\ \addlinespace[2pt] 
   
\multirow{2}{*}{\centering NB}   & \multirow{2}{*}{\centering 57.50}   &             & 60.54 &62.5 & 67.60  &              & 60.54  &62.51 &66.72  \\ \addlinespace[-2pt] 
   &  &                                                                               & (61.66) &(63.33) & (68.95)  &                         & (4.54)  &(4.41) &(4.13)  \\ \addlinespace[2pt]
   
\multirow{2}{*}{\centering DT}   & \multirow{2}{*}{\centering 73.33}  &              & 72.37 &71.08 & 72.06      &           & 72.04  &71.66 &70.72  \\ \addlinespace[-2pt] 
   &   &                                                                            & (73.95) &(74.58) & (73.54)  &            & (3.70)  &(4.38) &(4.10)  \\ \addlinespace[2pt] \hline \addlinespace[1pt]
\end{tabular}%
}
\parbox{\columnwidth}{\footnotesize Note: Values within parentheses under RHO are the standard deviation values and under 10-fold CV are the maximum values obtained.}
\end{table}

\subsection{Comparison with State-of-the-art (SOTA) Models}
\label{subsection:compare_SOTA}

The comparison with the SOTA models is given in Table~\ref{T:compare_SOTA}. Since the SF protocol is unique, it cannot be directly compared with previous works. However, WOBF and WBF protocols, which have been used in prior studies, were included in the comparison. These results are taken with 10-fold CV evaluation method, and the best accuracy under each model for each protocol is highlighted in bold. Additionally, since the train-test subsets vary across different studies, we compared these SOTA results with the maximum accuracy we achieved for each model under the 10-fold CV evaluation (mentioned in Table~\ref{T:compare_SOTA} as (Our)). Although our implementations underperformed with the WBF protocol, we achieved the best performance across all models with the WOBF protocol, showing accuracy margins of 10\%-20\%. These results indicate that while hyperparameters are transferable with simpler features (e.g., categorical features in WOBF), this is not the case with more complex features (e.g., numerical features in WBF). Furthermore, this comparison highlights the importance of standard evaluation protocols to ensure fair comparisons.

\begin{table}[tbp]
\centering
\caption{Accuracy comparison with SOTA models}
\label{T:compare_SOTA}
\resizebox{\columnwidth}{!}{
\begin{tabular}{lllllll}
\hline
\addlinespace[1pt]
Protocol                      & MLPC            & SVM             & k-NN            & LR              & NB              & DT              \\ \addlinespace[1pt] \hline
\addlinespace[1pt]
\multirow{5}{*}{WBF}  & \textbf{78.6}~\cite{rahman2017predict}   & \textbf{79.0}~\cite{roy2024adaptive}    & 73.0~\cite{roy2024adaptive}    & \textbf{78.0}~\cite{roy2024adaptive}    & 74.0~\cite{roy2024adaptive}    & 73.0~\cite{roy2024adaptive}    \\
                      & 79.1~\cite{amrieh2016mining}    & 64.2~\cite{francis2019predicting}   & \textbf{74.2}~\cite{rahman2017predict}   &                 & \textbf{74.0}~\cite{rahman2017predict}   & \textbf{77.5}~\cite{rahman2017predict}   \\
                      & 64.2~\cite{francis2019predicting}   &                 & 73.6~\cite{wafi2019automatic}    &                 & 67.7~\cite{amrieh2016mining}    & 75.8~\cite{amrieh2016mining}    \\
                      &                 &                 &                 &                 & 52.8~\cite{francis2019predicting}   & 66.0~\cite{francis2019predicting}   \\ \cline{2-7} \addlinespace[1pt]
                      & 78.1 {(}Our{)} & 77.3 {(}Our{)} & 67.1 {(}Our{)} & 75.8 {(}Our{)} & 63.3 {(}Our{)} & 74.6 {(}Our{)} \\ \addlinespace[1pt]\hline
\addlinespace[1pt]
\multirow{3}{*}{WOBF} & 57.0~\cite{amrieh2016mining}    &                 & 54.8~\cite{rahman2017predict}   &                 & 46.4~\cite{amrieh2016mining}    & 55.6~\cite{amrieh2016mining}    \\
                      & 55.3~\cite{rahman2017predict}   &                 &                 &                 & 54.1~\cite{rahman2017predict}   & 55.8~\cite{rahman2017predict}   \\\cline{2-7} \addlinespace[1pt]
                      & \textbf{70.8} {(}Our{)} & \textbf{70.2} {(}Our{)} & \textbf{63.8} {(}Our{)} & \textbf{69.6} {(}Our{)} & \textbf{69.0} {(}Our{)} & \textbf{73.5} {(}Our{)} \\ \addlinespace[1pt]\hline
\end{tabular}%

}
\end{table}

\subsection{Model Explanations}
\label{subsection:shap_res}
\begin{figure}[tbp] 
    \centering
    \includegraphics[width=\linewidth]{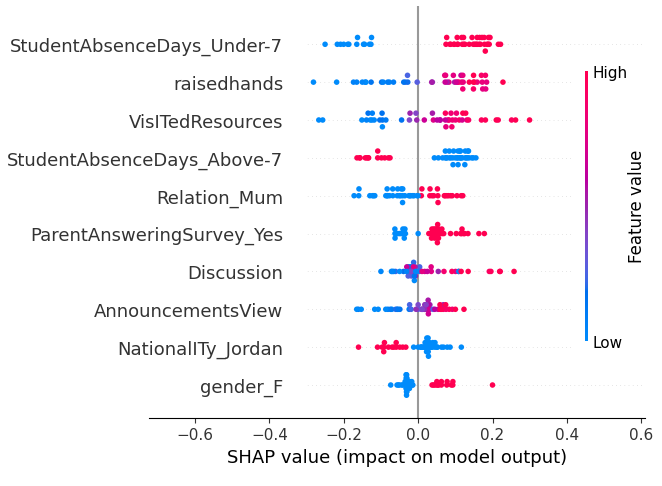}
    \caption{SHAP value distribution for 10 most impactful features in MLPC model (under 10-fold CV / SF protocols)}
    \label{fig:shap_mlpc_sf_10fold}
    \vspace{-3mm}
\end{figure}

SHAP value distribution of the MLP model is given in the Figure~\ref{fig:shap_mlpc_sf_10fold}. Out of the 48 samples of test set, 30 were correctly classified but the distribution in Figure~\ref{fig:shap_mlpc_sf_10fold} contains all samples. When considering the test set, we observe that student absence dominate the predictions across all samples. Moreover, all behavioral features are among the dominant features. When considering only the misclassified samples, student absence contribute is among the first 5 thus showing that student absent, while contribute to correctly predicting the output for, it also contribute to misclassifications. Furthermore, behavioral feature contribution is still present in the misclassified instances but to a lesser degree. Contrary to what was expected, academic features contribution is less intensive. This could be attributed to the large number of features resulted from  encoding process. Inclusion of all numerical features based on the EDA is jusified based on these SHAP distribution while it is further corroborated by the pearson coefficients we obtained for the features when calculated against the class label. Importance of student absence feature was noted in the EDA thus these findings corroborate it.

\section{Conclusion}
\label{sec:conclusion}
We have shown the importance of proper feature selection for model performance as evident by the higher average performance of FS across most of the models and evaluation protocols. Furthermore, MLP obtained 86.46\% maximum accuracy which is only second to maximum accuracy of SVM which is 88.65\%. Considering there were 1986 experiments, this suggest that neural networks can be used as a data-efficient machine learning model with SAPData dataset. Given the naive baseline is 37.08\%, performance of all models are acceptable while considering the relative performance, MLP comes 1st, SVM comes 2nd with DT coming in 3rd. When considering the stability of the models, DT behavior is invariance to evaluation protocol with low variance. Results from xML (SHAP) suggest that behavioral features correlate highly with the predictions in academic performance. While academic features correlation is less evicdent, this doesn't give any indications of causal effects. 

These results suggest several future research directions such as going beyond the MLP and exploring advanced neural network architectures such as CNN, LSTM,Transformers as well as data-efficient neural network architectures such as spiking neural networks (SNN), bayesian neural networks (BNN) and kolmogorov-arnold network (KAN). Moreover, improvements can be done to feature selection by combining feature selection with xML. Finally, implementations with ensemble, bagging and boosting methods can be done to obtain better benchmark results on SAPData dataset.



\bibliographystyle{IEEEtran}
\bibliography{references}

\end{document}